\documentclass[11pt]{article}
\usepackage{moriond,epsfig}
\usepackage{amsmath}
\usepackage{amsfonts}
\usepackage{amssymb}
\usepackage{graphics}
\usepackage{tabularx}

\def\be{\begin{equation}}
\def\ee{\end{equation}}
\def\bea{\begin{eqnarray}}
\def\eea{\end{eqnarray}}

\def\lsim{\:\raisebox{-0.5ex}{$\stackrel{\textstyle<}{\sim}$}\:}
\def\gsim{\:\raisebox{-0.5ex}{$\stackrel{\textstyle>}{\sim}$}\:}

\begin{document}
\vspace*{4cm}
\title{Unraveling flavor \& naturalness from RUN II to 100 TeV}\footnote{Invited
talk at the EW Moriond 2015}

\author{AMARJIT SONI}

\address{Department of Physics, Brookhaven National Laboratory,\\
 Upton, NY 11973, USA}

\maketitle
\abstracts{The importance of incorporating flavor constraints, when providing bounds on new physics is
stressed. As is well known it is very difficult for models of new physics to have scales lighter than about
10 TeV once flavor constraints are built in. Although, in the conventional sense, this higher scale means more
tuning, it may well make the underlying theory simpler as illustrated with one example.  Direct production
of new particles heavier than about 5 TeV becomes very difficult at LHC though  indirect signals
such as deviations in the Higgs to gamma gamma branching ratio may still be possible. Also
of course precision studies at low energy facilities can be useful for detecting new phenomena.
For direct production of new particles over 5-10 TeV a new high energy collider perhaps at 100 TeV is
highly desirable.  Indeed, a very strong physics case can be made for such a machine.  It can be used to significantly improve many bounds as well as allow a multitude of direct searches in many channels without relying on specific models. Naturalness can be tested
by another  factor of about 50 past LHC to O($10^{-4}$),  a valuable achievement in itself.
}  

\section{Introduction and Motivation}

In this talk I’d like to suggest (again~\cite{older}) that the expectations of new physics scale around 1-3 TeV were overly optimistic; they seem to have been deduced by only paying  attention to electroweak precision constraints[EWPC] and, as a rule,   experimental constraints from flavor physics were being disregarded.  Flavor physics constraints strongly suggest new physics scale is unlikely to be below $\approx$ 10 TeV.  
Studies in the past decade largely show that direct verification  of new physics heavier than around 5 TeV is highly unlikely to come  
from LHC. However, there is a tantalizing example of a precoscious signal 
from improved studies in RUN II of LHC of Higgs properties that may reveal that actually the 125 GeV
scalar is a Higgs-radion of Randall-Sundrm (RS)~\cite{GBS13,See_GW99}.  This object’s properties are very similar to that of the Standard Model Higgs [SMH] but not
exactly the same and indeed its BR into two gluons or two photons are appreciably bigger~\cite{GBS13}.

Importance of experimental constraints from flavor physics are emphasized; after all existence of flavors
is an experimental reality and the constraints resulting from them are profound statements of flavor
alignment. Bearing that in mind, it is suggested that estimates of new physics scale must obey constraints from flavor physics and not just EW precision constraints.  
This procedure very likely will   raise the scale of new physics and to that extent may require more tuning and therefore  the theoretical framework more fine-tuned.  However, as emphasizd before	~\cite{older}, this additional fine-tuning quite likely will be offset by a simpler underlying theory. In the explicit example of RS a lower and less tuned scale requires
the introduction of additional degrees of freedom  to ensure custodial symmetry and in fact with exotic charges~\cite{AMS03}.  However, this need not necessarily be the case.  The reason that it does not have to be that way is because it critically depends on how the underlying theory of flavor is constructed; it is plausible that when we have a genuine theory of flavor it will not be more tuned, as will be discussed.

Although the experimental constraints from flavor considerations can only give us a lower bound on the
scale of new physics, it is quite plausible that in order to keep the tuning minimal, the  scale of
new physics will actually be close to the flavor bound.

An illustrative  candidate theory of flavor is based on Randal-Sundrum warped space 
ideas~\cite{RS991,RS992}.  As is well known by now,  these ideas lead to an interesting geometric theory of flavor. In the underlying 5-dimensional construction the localization parameters  representing quark masses are all of
O(1)~\cite{MN09}. It leads to a natural RS-GIM mechanism\cite{APS04} so that FCNC in the light sector are significantly suppressed.
Large flavor violations, though, are expected in top \cite{APS07}; indeed large $t \to q Z(h), 
\gamma$ and gluon is one of
the many interesting generic predictions. Also O(1) BSM  CP-odd phases are expected based on naturalness arguments\cite{APS04,AJB08,MN08}.

However, closer studies show that FCNC have right-handed components. These lead to enhanced 
matrix elements and enhanced RG-running~\cite{BBS82,Bona07,LS09}, in particular in Kaon decays and in mixings. Consequently experimental constraints from
$\Delta m_K$ and $\epsilon_K$   constrain KK-gluon masses to be above $\approx$ 10 TeV.  This clearly represents a big success of an RS based theory of flavor as it significantly reduces the
tension between EW-Planck (around 1 TeV) hierarchy and (naive) flavor hierarchy from  (about 1000 TeV) by
about two orders of magnitude.

Despite some very attractive features, an RS theory of flavor currently also has several serious drawbacks, in particular in the lepton sector.  Moreover,  while shortcomings and problems with the SM are all widely recognized   such as neutrino masses, dark matter, baryogenesis,     unification etc                                                                             
, our attempts at making extensions of the SM, via RS, extra scalars or via SUSY  also face severe problems. There is an explosion of parameters, and in most instances there is no understanding of flavors, there are many unnatural aspects.
Therefore, it is highly plausible that      we are missing something basic when it comes to model building.     

For this reason as well,  a doze of experimental reality from a high energy collider of O(100 TeV) could do us wonders.  Given that naive flavor constraints tend to be around 1000 TeV, there is a very good chance that in a genuine theory of flavor (as e.g. in RS), the actual scale from flavor constraints would be a lot
smaller than 1000 TeV. Thus, 100 TeV collider may well unlock the mystery of flavor in some important ways.


\subsection{{\bf Higgs-radion interpretation of the 126 GeV scalar}}

On July 4th 2012, the LHC made the spectacular announcement of the discovery of a scalar with
mass around 126 GeV~\cite{ATLAS12,CMS12}. Since then studies of all its properties have so far shown that they
are all quite consistent with that of a Standard Model Higgs~\cite{GL13}. While this is yet another remarkable
achievement of the SM, it is important to remember that the accuracy of these measurements, at present,  is quite
limited and is only around 15-20\%; this means there is ample room for new physics to contribute
and we must continue making every effort to make these tests more precise.

\section{The pros and cons of an RS theory of flavor}

RS warped space ideas provide an elegant and  interesting resolution to the flavor puzzle. 
First recall that in the original RS model, the EW –Planck hierarchy is solved by localizing the Higgs
on the IR or TeV brane~\cite{RS991}.  Soon thereafter it was realized that these ideas based on the warped
metric also readily can be exploited to remove large hierarchies in masses of quark flavors by fermion geography and localization~\cite{GN00,GP00,DHR01}. Indeed the localization parameters characterizing the six quark masses
in the underlying 5-dimensional theory are explicitly all of O(1)~\cite{MN09}. There is a clear picturesque 
understanding of quark masses. Quarks localized close to the UV brane are light because they have less 
overlap with the Higgs which is localized on the TeV brane, whereas the quarks near the TeV brane
are heavy (like the top quark) because they have much greater overlap with the near by Higgs on the TeV brane.  
RS warped framework for flavor actually offers     
a  natural suppression        of   most  of the FCNC. This is especially impressive given that flavor-changing
transitions though occurring at the tree level,                                   resulting from rotations from interaction 
to mass basis, are suppressed roughly to the same level as the loop in the SM.  Thus the warped flavor construction has an automatic RS-GIM mechanism~\cite{APS04}.            Another interesting consequence
of this setup is that it predicts that     many BSM phases of O(1) occur quite naturally~\cite{APS04,APS04B,APS07,AJB08,MN08}.   Another
key generic prediction is that most flavor violations are driven by the top quark and in fact sizeable rates for                                                                                                                                                                                                                                                                                                                                                                                                                                                                                                                                                                                                                                       processes such as $t \to q Z [h, gluon, \gamma]$ occur rather naturally~\cite{APS07}
which are much bigger than in most BSMs.

It turns out that because the flavor-changing currents have right-handed components, they have enhanced effects on $K-\bar K$ mixings. This is for two reasons. First because of the larger matrix elements of LR currents and secondly because  of the enhanced RG-running of LR currents.       

Be that as it may, these enhancements have the implications that important physical observables  
such as experimentally measured value of $\Delta m_K$ and the indirect CP violation parameter, $\epsilon_K$
in $K \to \pi \pi$ provide a lower bound on KK-gluon mass of O(10 TeV).      	

This underscores the great success of an RS theory of flavor as          the fermion geography and simple localization ideas are able to lower the scale of new physics from quark-flavor sector from O(1000 TeV)
down to about O(10 TeV).                                


\section{Challenges of the lepton sector for a strictly geometric theory of flavor}

The key difficulty is that experimental searches for LFV have gotten very stringent over the years, see ~\cite{KKS14}
Table~\ref{table:constraints}.
 For example, the bound on
$\mu \to e \gamma$ now stands at $5\times 10^{-13}$ implying a new physics  scale over many tens of TeVs.  In fact all the bounds on channels that involve electron in the final state are very tight~\cite{KKS14}.    
In sharp contrast though, the muon (g-2) anomaly exhibits a very serious deviation of around 3.5 $\sigma$ from the SM. If this is to be interpreted in terms of new physics, the scale of new physics cannot be too high, most likely it should be below a TeV. Thus it may well be that the electron is a special case because of (possibly accidental) symmetry that it does not (effectively) participate in the LFV and the effective scale of new physics may well be not that high. Be that as it may, it is clear though that more work on RS type of theories of flavor especially for the lepton sector is needed.

\begin{table}[ht]
\centering
\begin{tabular}{| c | c |}
\hline
Observable & Limit \\ \hline \hline
Br($\mu\rightarrow 3e$) & $<1.0\times 10^{-12}$~\cite{PDG2014} \\ \hline
Br($\mu\rightarrow e\gamma$) & $< 5.7\times10^{-13}$~\cite{PDG2014} \\ \hline
Br($\tau\rightarrow 3e$) & $< 2.7\times 10^{-8}$~\cite{PDG2014} \\
Br($\tau\rightarrow e^-\mu^+\mu^-$) & $< 2.7\times 10^{-8}$~\cite{PDG2014} \\
Br($\tau\rightarrow e^+\mu^-\mu^-$) & $< 1.7\times 10^{-8}$~\cite{PDG2014} \\
Br($\tau\rightarrow \mu^-e^+e^-$) & $< .8\times 10^{-8}$~\cite{PDG2014}\\
Br($\tau\rightarrow \mu^+e^-e^-$) & $< 1.5\times 10^{-8}$~\cite{PDG2014} \\
Br($\tau\rightarrow 3\mu$) & $< 2.1\times 10^{-8}$~\cite{PDG2014} \\ \hline
Br($\tau\rightarrow \mu\gamma$) & $< 4.4\times 10^{-8}$~\cite{PDG2014} \\ 
Br($\tau\rightarrow e\gamma$) & $< 3.3\times 10^{-8}$~\cite{PDG2014} \\ \hline
$\mu-e$ conversion & $\Lambda \gtrsim 10^3$ TeV~\cite{deGouvea2013} \\ \hline 
$e^+e^- \rightarrow e^+ e^-$ & $\Lambda \gtrsim 5$ TeV~\cite{Schael2013} \\ 
$e^+e^- \rightarrow \mu^+ \mu^-$ & $\Lambda \gtrsim 5$ TeV~\cite{Schael2013} \\ 
$e^+e^- \rightarrow \tau^+ \tau^-$ & $\Lambda \gtrsim 4$ TeV~\cite{Schael2013} \\ \hline
\end{tabular}
\caption{Constraints on lepton-flavor violating and conserving processes. For the last four observables, the experimental null results are given in terms of a dimension-6 operator, suppressed by two orders of $\Lambda$, which can be interpreted as the nominal scale of new physics;  taken from ~\protect\cite{KKS14} }
\label{table:constraints}
\end{table}

\section{{\bf Experimental repercussions: From LHC RUN II to 100 TeV}}

As already emphasized there is an exciting possibility that LHC RUN II 
may reveal that the 125 GeV scalar is not a SM Higgs but rather a ``Higgs-radian''.  The distinguishing feature is that the Br into 2 gluons or 2 photons of Higgs-radian is about 50\% larger than that of the SM higgs; see Table~\ref{table:hr-Br} . Therefore, it is extremely important to measure these branching ratios as precisely as we can in RUN II. With the anticipated luminosity of  O(100/fb),  its quite plausible that these branching ratios will be measured well enough in the next few years, to confirm or rule out the Higgs-radian interpretation.

\begin{table}[htb]
\begin{center}
\begin{tabular}{|l||c|c|}
\hline
 & SM ($m_h=126~GeV$) & Higgs-Radion ($m_{h_r}=126~GeV$) \\
 \hline
$Br(h\to WW^*)$ & 0.231 &0.204 \\
$Br(h \to ZZ^*)$ & 0.0289 & 0.0257 \\
$Br(h \to gg)$ & 0.0848& 0.13\\
$Br(h \to \gamma\gamma)$ & $2.28 \cdot 10^{-3}$ & $3.8\cdot 10^{-3}$ \\
$Br(h \to b \bar b)$ & 0.561 &  0.545 \\
$Br(h \to \tau \bar\tau)$ & 0.0615 & 0.063\\
$Br(h \to c \bar c)$ & 0.0283 & 0.028 \\
Total width [GeV] & $4.21 \cdot 10^{-3}$ &  $2.2 \cdot 10^{-3}$ \\
\hline
\end{tabular}
\caption{The Higgs-radion and the SM Higgs branching ratios and total width. The SM
values are taken from~\protect \cite{Higgs-handbook}; note the  effect of KK-tower is not included.This table is taken from ~\protect\cite{GBS13}}
\label{table:hr-Br}
\end{center}
\end{table}

\begin{figure} [htb]
\begin{center}
\includegraphics[width=0.7\hsize,angle=90]{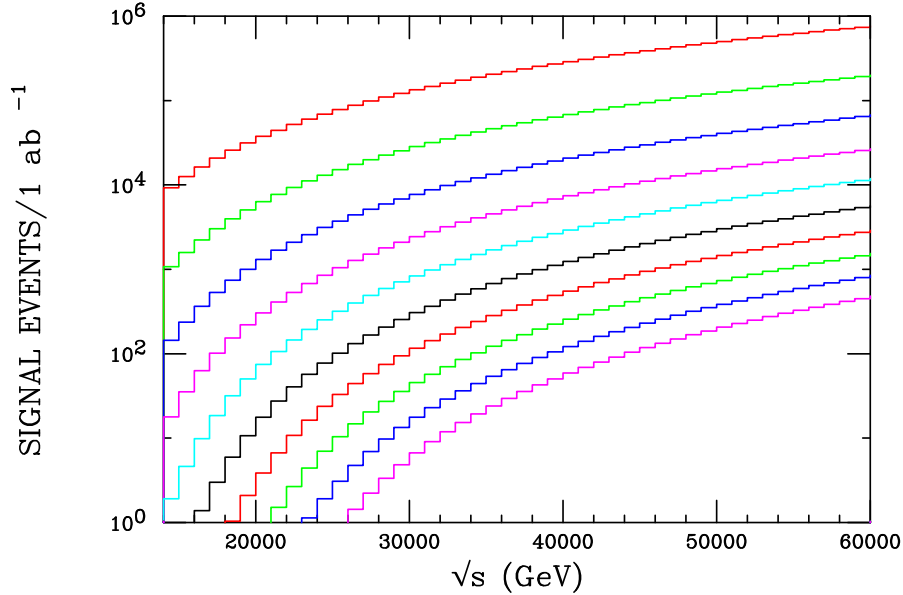}
\caption{Expected rate for KK gluon $\to$ top pairs 
is shown versus collider energy for KK glu masses from
3 to 12 TeV. Branching fractions and efficiencies
are neglected;  taken from ~\protect\cite{DRS07} where more information on cuts and other details may be found}
\label{gluKKreach_S}
\end{center}
\end{figure}

Using the measured properties of the 125 GeV scalar~\cite{GL13} , it was shown that if the Higgs-radian interpretation of the data is viable, then the KK-gluon mass in the underlying RS model is around 4.5 TeV~\cite{GBS13}. Direct detection of a KK-gluon of mass around 4.5 TeV at LHC energies is likely to be extremely challenging~\cite{AP_kkg08}; higher energy collider will be needed; see
 fig.~\ref{gluKKreach_S}.
Indeed, as discussed before, flavor constraints seem to require KK-gluon needs to have a mass over about 10 TeV.
This provides an even stronger reason for a higher energy collider.

Actually, even if the KK-gluon mass is around 4.5 TeV, the KK-graviton 
mass is expected to be about 40\%    larger {\it i.e.} about 6.3 TeV~\cite{ADPS07} . Unfortunately, at LHC it is virtually impossible  to directly detect gravitons of mass larger than about 3 TeV.~\cite{ADPS07,LR07,KA13_snowmass}  The KK-graviton is a unique and distinctive particle and a hallmark of RS theories and its production and study is highly motivated and for that purpose too a higher energy collider is quite essential~\cite{DRS07}.

In fact the physics case for a higher energy collider such as 100 TeV is pretty compelling with or without RS. We have to acknowledge that the LHC RUN I with no sign of new physics has turned  the field in somewhat of a disarray.  While it is quite easy for any of us to find shortcomings of the SM, the BSMs that are being proposed all seem to have even more serious problems. They all look a lot more complicated with many more parameters than in the SM. Most of the models give no understanding of flavors or masses. Given the disarray and confusion that has become the norm,  a dose of experimental reality from a new high energy collider could be extremely  useful. 

Note also that 100 TeV has some special significance.  First of all we should recall that over 25 years ago, a machine with center of mass energy of 40 TeV called the SSC (Superconducting Super Collider) was being seriously planned and built, near Dallas, Texas.    After couple of years of digging a substantial fraction of the tunnels at a cost of a few billions of dollars, the project was cancelled. This resulted mostly from failure of a political system rather than any perceived technical difficulty.

{\bf What is so special about 100 TeV scale?}

As is well known, we are faced with two famous puzzles in Particle Physics: the Electroweak- Planck  hierarchy and the flavor puzzle.
The first refers to the fact that if we assume there is new physics beyond the SM at a scale $\Lambda$, the higgs self-energy tends to receive a correction of O($\Lambda/m_h)^2$ which implies that 
to avoid excessive fine-tuning,  $\Lambda$ needs to be below about a TeV. On the other hand any generic new physics tends to cause large
FCNC and entities such as $\Delta m_K$ and $\epsilon_K$ require $\Lambda_{flavor} \gsim 10^3$ TeV.  In an ideal model of new physics and flavor $\Lambda \approx \Lambda_{flavor} \lsim TeV$. What makes RS an interesting and compelling theory of flavor is that it tends  to give $\Lambda \approx$ 3 TeV and $\Lambda_{flavor}\approx$ 10 TeV
in so far as FCNC in the quark sector are concerned; the difference is only a factor of O(3). It is therefore not
inconceivable that in a genuine theory of new physics and quark and lepton flavor, the relevant mass scale is not a lot bigger than about 10 TeV.  It is therefore quite plausible that a 100 TeV collider could   reveal vital information on the underpinnings of flavor.

\subsection{{\bf Huge physics opportunities: Its a Goldmine!; Its a No-Brainer!}} 

\begin{description}

\item A hadron collider with ~100 TeV is a powerful discovery machine.
First of all there is significant increase in the production rates
over LHC for many interesting and important final states; see  fig.~\ref{qcd-xsecs-mcfm-Edep} ~\cite{campbell_snowmass}; as one important example we can expect over $10^{10}$ top pairs\!.

\begin{figure}[htb]
\begin{center}
\includegraphics[width=0.7\hsize,angle=90]{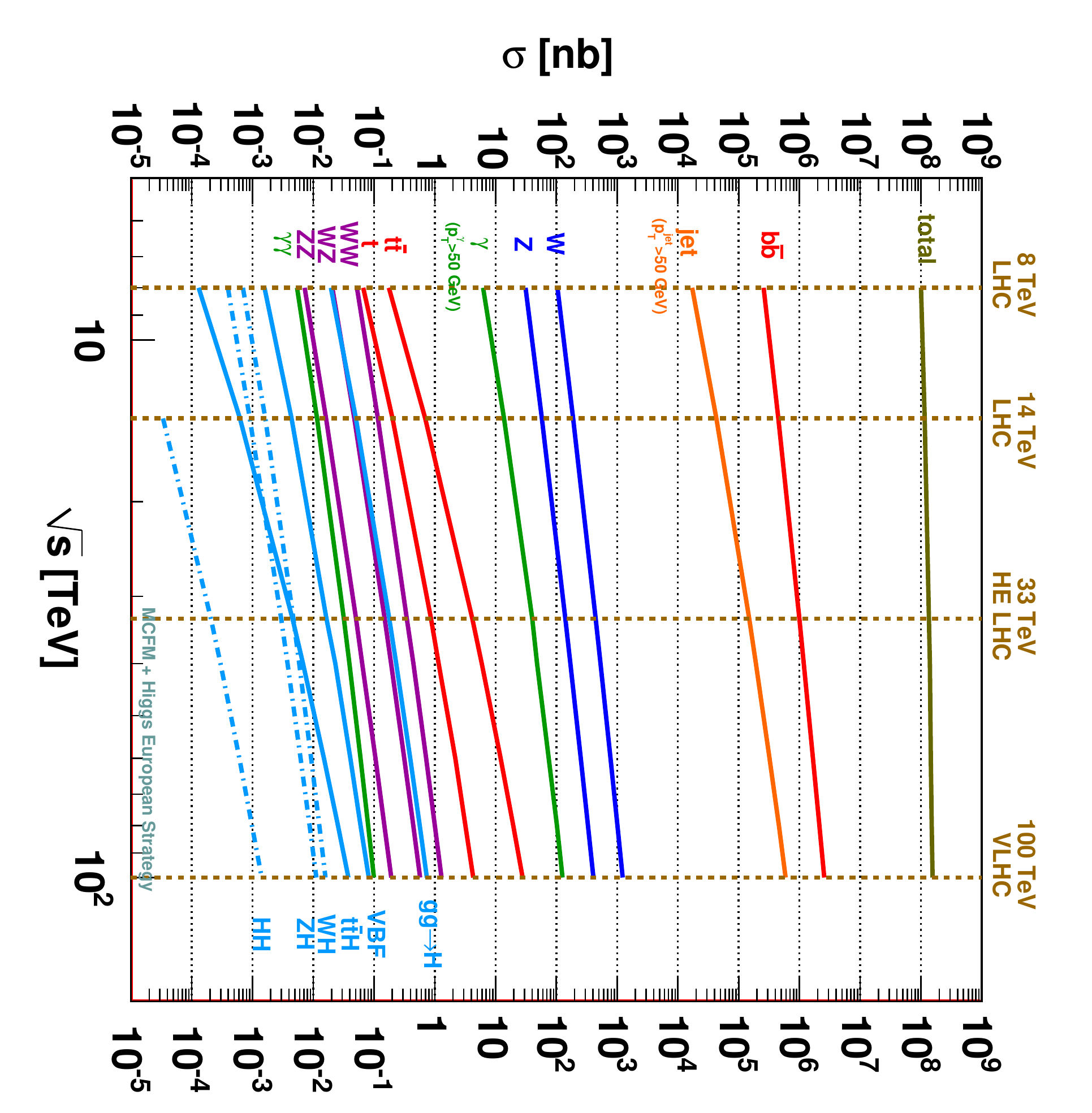}
\caption{Cross section predictions at proton-proton colliders
as a function of center-of-mass operating energy, $\sqrt s$; taken from ~\protect\cite{campbell_snowmass}.
}
\label{qcd-xsecs-mcfm-Edep}
\end{center}
\end{figure}

\item  Therefore, this machine will  allow us to significantly improve many bounds that are of crucial importance.

Below are some examples:

 a) Flavor changing decays of the top,  t $\to$ q h, q Z
where q=c or u~\cite{t_fcnc}.

b) $t_{partner} \to$ t h, relevant to higgs radiative stability.

 c) Same-sign top quark pairs, a powerful diagnostic test of new physics\cite{AGS_SST,Berger11,AguilaSaavedra11,Degrande11}.

d) h, H, Z, Z' $\to \mu + \tau$  as, for example, studies in~\cite{BHSS85,TLS10}

e)  h $\to$ Z, $Z^{\prime} \to 4$ leptons, gold-plated final state  for in-depth study in search of new physics with or w/o CP 
violation~\cite{XuS95,Gainer13,Bohdan13,Berge14,Chen14,Chen15}

f) t $\to$ q $\tau \mu$ in search of LFV~\cite{KS08}

g) $t \bar t h$ final state is very important  for searching  signs of new physics in CP violation~\cite{SBS95}
and many other observables.

h) The  popular RS models predict a zoo of new particles: KK-gluon, graviton, Z, W, fermions.......
At a 100 TeV collider the reach for these particles  goes  to about 15 TeV~\cite{DRS07}.  

The reach for right-handed gauge bosons, relevant to the case of the interesting Left-Right Symmetric models  can also be extended to about 15 TeV. Just as we have a ``generational puzzle'' , embodied in Rabi's legendary phrase,
``Who ordered the muon'', it is not inconceivable  that the Higgs doublet of the SM may repeat  itself.
At a 100 TeV collider searches for the 2nd Higgs doublet can be extended to a few tens of TeVs.


i) LHC should allow us to directly search for many types of new physics upto a mass scale of around around 5 TeV.  In terms of tuning this
translates to about O($(v \approx 0.25 TeV/5 TeV)^2$)  $\approx$ O($ few \times 10^{-3}$). In going from LHC to 100 TeV, fine tuning  will be probed further by another factor of $\approx 50$, {\it i.e} to about $ few \times 10^{-5}$. This is an important achievement in itself.

\end{description}


\section{{\bf Experimental consequences for the low energy, precision (intensity) frontier assuming the scale
of new physics is $\approx$ 10 TeV}}.

Going forward, as always, a multi prong attack in search of new physics is essential. 
Below is a list of some of the most important low energy probes.

\begin{description}

\item Searching for neutron electric dipole moment (nedm)  is extremely important as in a large class of new physics models 
it is orders of magnitude bigger than in the SM~\cite{MRM13}

\item Time-dependent CP in exclusive radiative B-decays. It was shown~\cite{ags,aghs} that time-dependent CP asymmetries in
modes such as B, $B_s \to P_1 P_2 \gamma$, (where $P_1$, $P_2$,  are pseudoscalars)  in the SM  must be vanishingly small and this represents a very clean null
test of the SM.  For  BELLE II and  LHCb this represents a very important means for searching new physics.

\item Precise determination of the angle $\gamma$ of the unitarity triangle. From a theoretical standpoint, given large enough
data samples $\gamma$ can be extracted exceedingly precisely~\cite{BZ13}.  Remarkably the underlying methods are completely data
driven~\cite{gamma_methods}. It is important therefore to use this opportunity as best we can to test the consistency of the SM.

\item From rare K-decays, precise measurement of the extremely challenging $K_L \to \pi^0 \nu \bar \nu$~\cite{LL89} remains a very important goal.
As is well known, in the SM this observable is really pristine; as was the previous example. This is why experimental measurements
though very challenging deserve a high priority~\cite{Shiomi14}

\item Precise calculation of the direct CP violation parameter, $\epsilon^{\prime}/\epsilon$ in the SM is very important
to see if it agrees with the measured value~\cite{PDG14} of $(16.6 \pm 2.3   )\times 10^{-4}$. Indeed after decades of relentless efforts, just recently,
the first lattice calculation ~\cite{RBC-UKQCD15} of this important parameter was completed with controlled errors, yielding the value $(1.38 \pm5.15  \pm 4.43) \times 10^{-4}$ which is consistent with the experimental number  within about 2 $\sigma$.  This is the 1st clear demonstration that the calculation is amenable to lattice techniques. Efforts are underway for improved calculations and there are good reasons to believe that in a few years time an improved calculation with total error of around 10\% of the experimental value can be attained~\cite{RBC-UKQCD15}.

Recall that theoretically reliable calculations of direct CP-asymmetries remain an important outstanding challenge.
In fact that in the B-system large direct CP asymmetries may arise from the KM mechanism~\cite{KM73} was anticipated
decades ago~\cite{BSS79}.  First experimental confirmation came fom BaBar and BELLE some 25 years later.
As a concrete example, direct CP asymmetry in~\cite{PDG14} $B^0 \to K^+ \pi^- = -0.082 \pm 0.006$
{\it i.e.} well over $10^4$  times bigger than $\epsilon^{\prime}$. To this day no reliable theoretical
calculation of this asymmetry exists and the possibility that a part of this may well be due
BSM-physics cannot be ruled out.

\subsection{{\bf Possible implications for {\bf {\it naturalness}} of precision low energy studies}}

Many of the tests involving CP-violating observables are exceedingy sensitive to BSM-CP phases. It is important to realize that naturalness arguements suggest that most BSM scenarios will entail new CP-odd phases. As we continue improving the experimental searches {\it e.g.} for nedm, time-dependent 
CP-violation in exclusive radiative B, Bs decays and ruling out the possibility of new physics in $\epsilon^{\prime}/\epsilon$ by significantly improving the
theoretical calculations then the ensuing null results will continue to push higher the scale of new physics thereby testing naturalness.

                                                                                                                                                                                                                                                                                                                                                                                                                                                                                                                                                                                                                                                                                                                                                                                                                                                                                                                                                                                                                                                                                                                                                                                                                            
\end{description}

%
%


\section{Summary}

It is emphasized that in projecting the scale of new physics, constraints from flavor physics should
be incorporated. Once we do that it is difficult to make a strong case for the scale of new physics
to be below about 5 TeV.  An interesting exception to this is pointed out which occurs
in RS type models wherein it is found that the 125 GeV scalar particle may exhibit appreciable
deviations in $\gamma$ $\gamma$ and glue glue final states from the predictions of the SM.
An important consequence of this suggestion regarding adherance of flavor constraints is
that it readily explains or accounts for the absence of any signals of new physics so far at the LHC.
Indeed, if the scale of new phenomena is 5 TeV or higher, then unfortunately, its
unlikely at LHC energies such heavy states will be produced. Indirect signals like deviation
in the branching ratio(s) of Higgs decay may still be possible. Also low energy precision
tests such as nedm and/or $\epsilon^{\prime}$ or flavor/CP studies in facilities such as LHCb or BELLE II
may bear fruit too.

The higher scale of new physics that results from inorporating flavor constraints
will of course entail more tuning in the sense that [Electroweak scale/new physics scale] becomes
smaller; however,  it is likely to make the underlying theoretical construction simpler as
illustrated.

Higher energy collider say at 100 TeV also is very attractive. It will allow us to put
bounds on multitude of observables and look for new phenomena in many channels 
with scales up to about 20-30 TeV. This scale is high enough that at least in
some interesting models even flavor constraints can be bypassed leaving open the possibility of
seeing some exciting signal(s).

At such scales even avenues in lepton flavor violation may have a chance
of spectacular observable effects as existing bounds from low energy tests may be surpassed.

Given our lack of success in building simple models for  extensions of the SM,
having a hadron collider at such high energies will give us a chance to cast a very wide net and search
in a very large class of final states. Thus, we do not need to rely on specific
models. This may be critical since current BSM models do not seem elegant
especially from the point of view of number of parameters.

Lastly, such a machine will allow us to extend our limits on fine-tuning and naturalness
by another factor of about 50  to the level of around $10^{-4}$, an important goal
in itself.

\section*{Acknowledgements}

Also I must  thank the organizers of the EW Moriond 2015, and in
particular Jean-Marie Frere for inviting me.
This work is supported in part by the US DOE contract No.
DE-AC02-98CH10886.


\begin{thebibliography}{99}
\bibitem{See_GW99}
The higgs-radion in our~\cite{GBS13} formulation is quite distinct from the one in the conventional Goldberger-Wise
mchanism~\cite{GW99}.
\end{thebibliography}
\end{document}